\documentclass[conference]{IEEEtran}
\ifCLASSINFOpdf
\else
\fi
\usepackage{amsmath}
\usepackage{amssymb}
\usepackage{bm}
\usepackage{graphicx}
\usepackage{algpseudocode}
\begin{document}
%
\title{INFusion: Diffusion Regularized Implicit Neural Representations for 2D and 3D Accelerated MRI Reconstruction}


\author{
    \IEEEauthorblockN{Yamin Arefeen\IEEEauthorrefmark{1}\IEEEauthorrefmark{2}, Brett Levac\IEEEauthorrefmark{1}, Zach Stoebner\IEEEauthorrefmark{1}, Jonathan I Tamir\IEEEauthorrefmark{1}}
    \IEEEauthorblockA{\IEEEauthorrefmark{1}The University of Texas at Austin \\ Chandra Family Department of Electrical and Computer Engineering}
    \IEEEauthorblockA{\IEEEauthorrefmark{2}MD Anderson Cancer Center\\ Department of Imaging Physics}
}


%


\maketitle

\begin{abstract}
Implicit Neural Representations (INRs) are a learning-based approach to accelerate Magnetic Resonance Imaging (MRI) acquisitions, particularly in scan-specific settings when only data from the under-sampled scan itself are available. Previous work demonstrates that INRs improve rapid MRI through inherent regularization imposed by neural network architectures.  Typically parameterized by fully-connected neural networks, INRs support continuous image representations by taking a physical coordinate location as input and outputting the intensity at that coordinate.  Previous work has applied unlearned regularization priors during INR training and have been limited to 2D or low-resolution 3D acquisitions.  Meanwhile, diffusion-based generative models have received recent attention as they learn powerful image priors decoupled from the measurement model.  This work proposes INFusion, a technique that regularizes the optimization of INRs from under-sampled MR measurements with pre-trained diffusion models for improved image reconstruction. In addition, we propose a hybrid 3D approach with our diffusion regularization that enables INR application on large-scale 3D MR datasets.  2D experiments demonstrate improved INR training with our proposed diffusion regularization, and 3D experiments demonstrate feasibility of INR training with diffusion regularization on 3D matrix sizes of $256\times256\times80$.
\end{abstract}


%
\IEEEpeerreviewmaketitle

\section{Introduction}

Magnetic Resonance Imaging (MRI) is a clinically important medical imaging modality that provides excellent soft tissue contrast without any ionizing radiation. However, MRI acquisitions suffer from an inherent trade-off between scan time, resolution, and signal-to-noise ratio (SNR) \cite{nishimura}. Shortening MRI scan time while maintaining image quality through computation reduces costs to patients and hospitals and reduces sensitivity to motion artifacts.

MRI scans can be accelerated by sampling the acquired Fourier (k-space) data below the Nyquist sampling rate and solving the ill-posed reconstruction problem to produce a high fidelity image from under-sampled measurements. To reduce scan time, clinics routinely use parallel imaging, \cite{grappa, sense,espirit}, which interpolates missing data with the encoding capability of the multi-coil signal receive array, and compressed sensing \cite{lustig-cs}, which exploits random sampling and sparsity. Recently, machine learning algorithms for accelerated MRI reconstruction employing training databases yield state of the art results \cite{modl,varnet,ssdu,robustcs,energy,baymri,mlreview}. End-to-end methods learn a point-wise mapping between under-sampled and fully-sampled data but are susceptible to test time shifts in the measurement operator. More recently, generative methods have been proposed that learn a prior over clean images and are robust to test time shifts in the forward operator.

All the MRI reconstruction algorithms previously discussed assume that a discretized grid of voxels represents the desired MRI image. However, alternative representations can also be employed, such as Implicit Neural Representations (INRs) \cite{siren,fourierfeatures}. An INR trains a neural network to take as input a spatial coordinate and output the signal value of interest at that spatial coordinate, thus providing a continuous image representation, and in some sense, implicitly storing the image in the weights of the neural network. Recent work utilizes INRs for accelerating MRI, particularly in scan-specific settings where only data from the under-sampled scan itself are available. By combining traditional regularization with the inherent regularization of the dense neural network architecture, positional encoding, and periodic activations, previous work demonstrates that INRs help further accelerate MRI scans \cite{imjsense,nerp,inrcardiacbin,inrfouriermri,nerfmri}. However, these methods apply unlearned spatial regularization priors and are mainly limited to 2D or low-resolution 3D acquisitions.

This work proposes further accelerating MRI with INR-based reconstructions by drawing inspiration from recent work in computer vision that uses diffusion priors to reconstruct INR-representations of 3D scenes from few camera views \cite{reconfusion}. In particular, our method, termed INFusion, uses diffusion models pre-trained on MRI images to regularize the optimization of INRs from under-sampled measurements. First, we demonstrate that our proposed approach improves reconstruction performance in comparison to competing techniques in 2D, $T_2$-weighted brain MRI. Then, we propose a hybrid 3D approach with our diffusion regularization that enables INR application on large-scale 3D MR datasets with 2D diffusion models, and we show experiments on a realistic, 3D knee acquisition.

\section{Methods}
\subsection{The Accelerated MRI Inverse Problem}
Given the linear, multi-channel MRI measurement model $A = P F S \in \mathbb{C}^{(M C) \times N}$, solving the following inverse problem yields an image from under-sampled measurements, $y\in\mathbb{C}^{M C}$, acquired in an accelerated scan, 
\begin{equation}
\arg\min_x ||y-Ax||_2^2 + \lambda R(x),
\end{equation}
where $x \in \mathbb{C}^{N}$ is the discretized image to reconstruct, $S \in \mathbb{C}^{(N C) \times N}$ describes how the multi-coil receive array measures signal, $F \in \mathbb{C}^{(N C) \times (N C)}$ applies the Fourier Transform to each coil image, $P \in \mathbb{C}^{(M C) \times (N C)}$ indicates which k-space points are sampled (with $M<N$ for reduced scan time), and $\lambda R(\cdot): \mathbb{C}^{N} \rightarrow \mathbb{R}$ imposes regularization (e.g., sparsity, low-rank).  

\subsection{Implicit Neural Representations}
Let $r \in \mathbb{R}^{3}$ represent a physical coordinate corresponding to a spatial location of the MRI anatomy to image (without loss of generality assume that $r \in [0,1]^{3}$). Then, an Implicit Neural Representation (INR) is a fully connected neural network $I_{\theta}: \mathbb{R}^{3} \rightarrow \mathbb{C}$ that takes a spatial coordinate, $r$, as input and outputs the signal value of interest of the MRI anatomy at that spatial location. Given a discretized image $x$, the INR can be trained by optimizing
\begin{equation}
\theta^* = \arg\min_{\theta} ||x - I_{\theta}(\pmb{r})||_2^2,
\end{equation}
where $\pmb{r} \in \mathbb{R}^{3 \times N}$ is the spatial coordinates of the voxel locations in the discretized image. Then, $I_{\theta^*}$ can be evaluated at spatial coordinates not on the original grid of coordinates, $\pmb{r}$, thus yielding a continuous representation.

\subsection{Accelerated MRI with INRs}
Given the discrete, under-sampled k-space measurements $y$, the following optimization problem estimates weights of an INR in the accelerated MRI setting:
\begin{equation}
\theta^* = \arg\min_{\theta} ||y - A I_{\theta}(\pmb{r})||_2^2 + \lambda R(I_{\theta}(\pmb{r})),
\end{equation}
recalling that $\pmb{r}$ is the spatial coordinates of the discretized voxel locations. Then evaluating $I_{\theta^*}(\pmb{r})$ yields the desired image. Previous work shows that the structure of the INR helps regularize the ill-posed reconstruction problem and yields higher fidelity images in comparison to directly solving for the discretized image \cite{imjsense,nerp,inrcardiacbin,inrfouriermri,nerfmri}. However, previous work regularizes the optimization problem with a hand-crafted spatial prior $R$, leaving room for further improvements with learned regularizors.

\subsection{INFusion: Regularizing INRs with Diffusion Models}
Our proposed INFusion method regularizes the INR MRI inverse problem with pre-trained generative diffusion models as they are powerful priors that decouple from the measurement model \cite{robustcs,energy,baymri} enabling applications with INRs. As it is computationally expensive to evaluate the diffusion model at every 3D coordinate, we also propose a stochastic regularizer in which the diffusion model is queried at random spatial coordinates $z \subset r$ in each iteration.

\begin{figure}[!t]
    \centering
    \includegraphics[width=\linewidth]{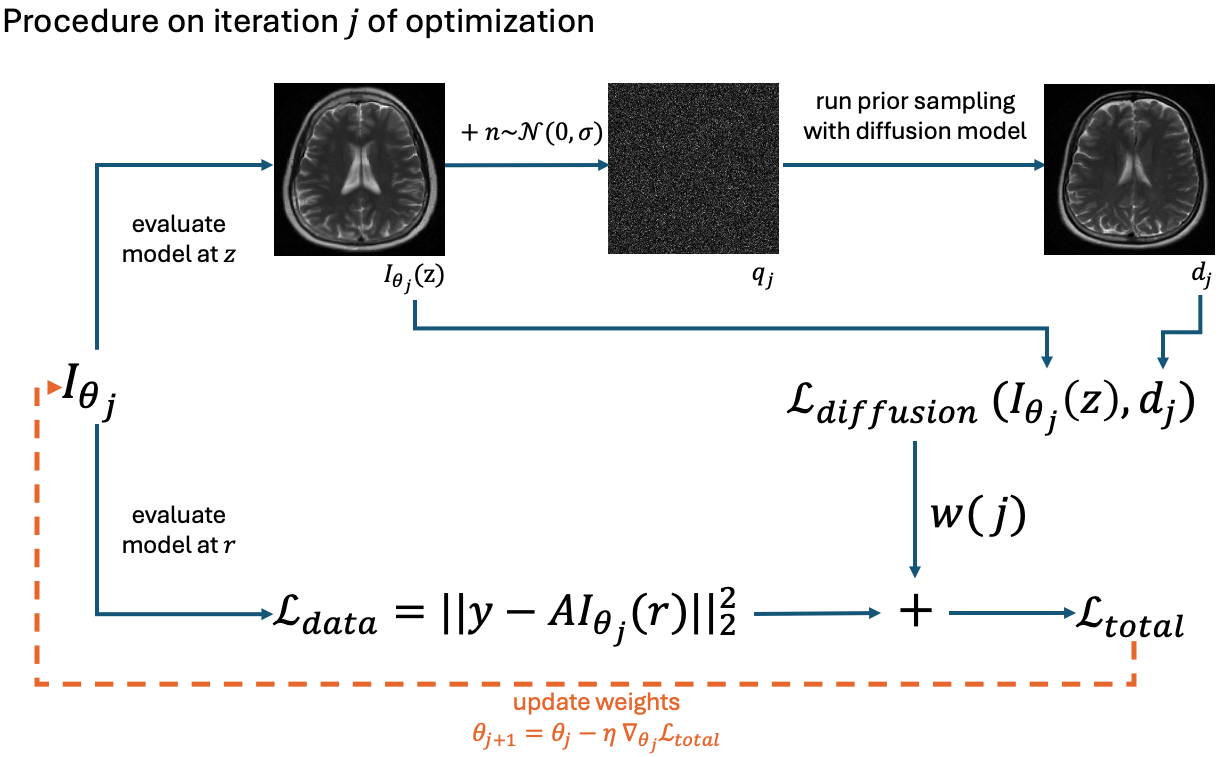}
    \caption{Our proposed approach, INFusion, first computes standard data consistency loss $L_{data}$ by evaluating the INR at discrete voxel locations, applying the MRI forward model to the resultant image, and then comparing to the acquired k-space data.  A second loss $L_{diffusion}$ is computed by adding gaussian noise to random slices produced by the INR, running prior sampling with the diffusion model, and comparing the resultant prior sample to the INR produced random slices with perceptual loss.  The diffusion loss guides the INR to produce images that better match the learned prior.}
    \label{fig:schematic}
\end{figure}

During each iteration, $j$, of optimization, we propose the following procedure:
\begin{algorithmic}[1]
\State Compute data consistency loss, $L_{data} = ||y - A I_{\theta_j}(r)||_2^2$.
\State Select spatial coordinates $z$ so that evaluating $I_{\theta_j}(z)$ yields random discretized 2D slices from the current estimate of image volume $x$.
\State Compute $q_j = I_{\theta_j}(z) + n$ where $n \sim \mathcal{N}(0,\sigma^2)$ and $\sigma \sim \mathcal{U}(\sigma_{\min},1)$.
\State Run diffusion \cite{stochasticscore,edm} with initialization $q_j$ to give a prior sample $d_j$.
\State Compute $L_{diffusion}=LPIPS(I_{\theta_j}(z),d_j)$ with LPIPS perceptual loss \cite{lpips}.
\State Compute $L_{total}=L_{data} + w(j) L_{diffusion}$ with adaptive scalar weighting $w(j)$, and backpropagate to update $\theta_j$.
\end{algorithmic}
Fig. \ref{fig:schematic} presents a schematic summary of the proposed procedure. In 3D settings, evaluating the INR at all spatial coordinates and backpropagating through the prior sampling procedure becomes computationally prohibitive. For 3D problems, we propose only encoding the $x-y$ dimension with spatial coordinates and having the model return discretized values corresponding to the third $z$ dimension. In addition, we compute our proposed diffusion regularization with respect to two random slices at each iteration.

\subsection{INR and INFusion Hyperparameters}
For all experiments, our INRs consisted of fully-connected neural networks with relu activations and 4 hidden layers using spatial coordinates encoded with 128 gaussian fourier features \cite{fourierfeatures}. INRs for 2D and 3D experiments had size 256 and size 3200 hidden layers respectively.

For the INFusion procedure, we selected $\sigma_{\min}$ to decrease from [1,0] linearly as function of the current optimization iteration $j$. Similarly, $w(j)$ decreased exponentially from [1,0]. We ran 10 iterations of the diffusion sampling formulation in \cite{edm} to generate the prior sample $d_j$.

\subsection{Generative Diffusion Model Training}
We trained two diffusion models using a U-Net style architecture with 65 million parameters and used the "EDM" hyper-parameters, loss function, and diffusion noise schedule \cite{edm}. The first model was trained on 8500 $T_2$ axial brain slices from fastMRI \cite{fastmri} of dimension $192 \times 192$ and the second was trained on 15720 sagittal knee slices from SKM-TEA \cite{skmtea}.

\section{Experiments}
\subsection{In-vivo 2D Single- and Multi- Coil Brain}
We used axial, T2-weighted, single-coil and 4-coil 2D k-space from the fastMRI \cite{fastmri} dataset, cropped both to $192 \times 192$ resolution, and retrospectively under-sampled with a 2D-Poisson pattern generated with BART \cite{bart} corresponding to an acceleration rate of $R=4$ for single-coil ($R$ times less than Nyquist) and $R=6$ for multi-coil. Both experiments compared the discretized image produced by solving the standard MRI inverse problem with L1-Wavelet regularization \cite{csreview} to an INR trained with (i) none, (ii) L1-Wavelet, or (iii) our proposed diffusion regularization. 

Second, we took 96 additional, multi-coil 2D samples from the fastMRI dataset, and retrospectively under-sampled them with 2D-Poisson under-sampling masks with an acceleration rate of $R=8$ and $R=9$.  These samples had coil counts ranging from $4 - 20$.  We compared error of the discretized images produced by solving the standard MRI inverse problem with L1-Wavelet regularization and INRs with L1-Wavelet and our proposed diffusion regularization.

\subsection{In-vivo small 3D Single Coil Brain and large 3D Multi-coil Knee}
We used a multi-slice single-coil $192 \times 192 \times 16$ dataset from fastMRI with axial brain slices, and treated it as 3D by retrospectively under-sampling it in the $y-z$ direction with a $R=2$ Poisson under-sampling mask. We also used k-space of an 8-coil 3D knee volume from the SKM-TEA \cite{skmtea} dataset, resized it to $256 \times 256 \times 80$, and uniformly under-sampled by $R = 4 \times 1$ in the $y-z$ directions. Both 3D experiments compared discretized images produced by an INR trained with and without our INFusion method. The smaller 3D brain experiment encoded all three spatial dimensions with coordinates and computed diffusion regularization across all 16 slices each iteration. The larger knee experiment incorporated the previously discussed modifications for 3D computational feasibility. Note that our diffusion model was trained on $x-y$ images, but under-sampling was performed in the $y-z$ direction.

\section{Results}
\begin{figure}[!h]
    \centering
    \includegraphics[width=\columnwidth]{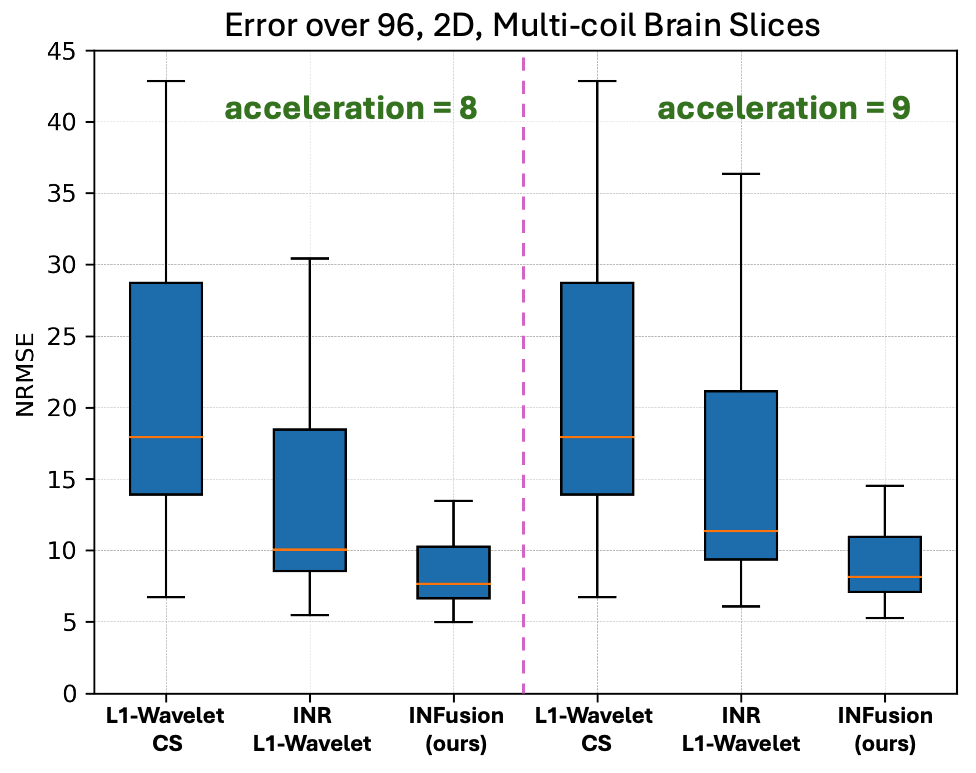}
    \caption{Quantitative comparison of standard L1-Wavelet CS and INRs trained with Wavelet and our proposed regularization at $R = 8$ and $R = 9$ across 96 samples.  Our proposed INFusion approach improves NRMSE across the test dataset.}
    \label{fig:2dresults-samples}
\end{figure}
\begin{figure*}[!t]
    \centering
    \includegraphics[scale=.55]{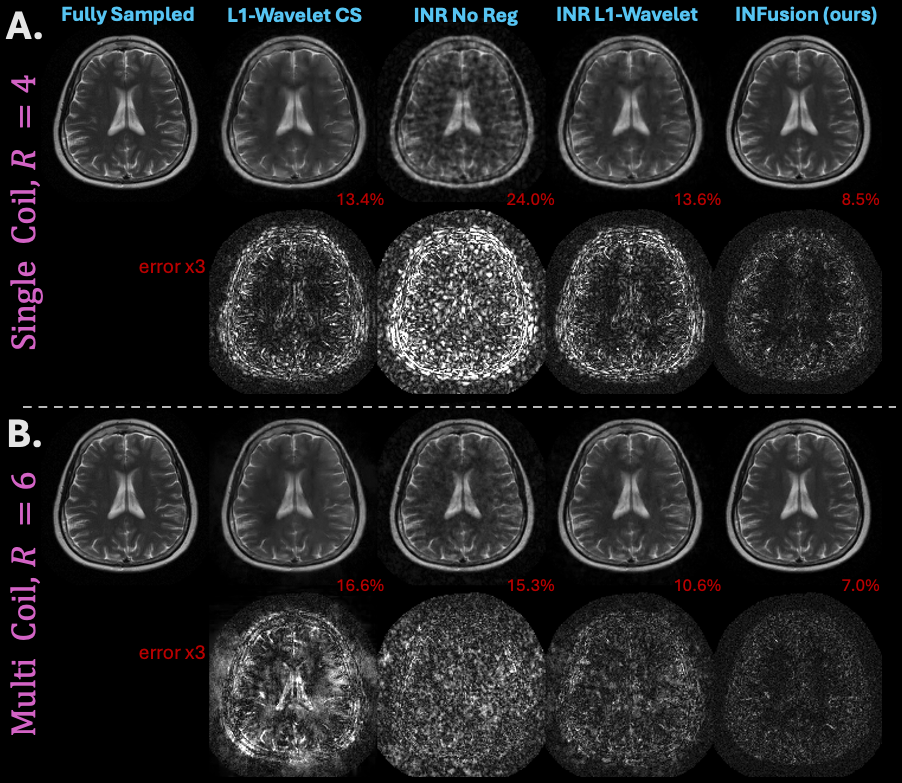}
    \caption{Reconstructions of discrete images on the (A.) 2D single-coil brain data at $R=4$ and (B.) 2D multi-coil brain data at $R=6$ with standard L1-wavelet Compressed Sensing (CS) and INRs with none, L1-wavelet, and our proposed diffusion regularization.  The proposed INFusion approach yields images with lowest NRMSE and best qualitative image quality.}
    \label{fig:2dresults}
\end{figure*}

INFusion achieved lower normalized-root-mean-squared-error (NRMSE) on the single (Fig. \ref{fig:2dresults}, A) and multi (Fig. \ref{fig:2dresults}, B) coil brain slice in comparison to INRs trained with Wavelet or without regularization and the standard Compressed Sensing (CS) L1-Wavelet MRI inverse problem.  The box plot in Fig. \ref{fig:2dresults-samples} shows that INFusion improves reconstruction performance with respect NRMSE over 96 slices in comparison to standard CS L1-Wavelet and INRs with Wavelet regularization.

\begin{figure*}[!t]
    \centering
    \includegraphics[scale=.385]{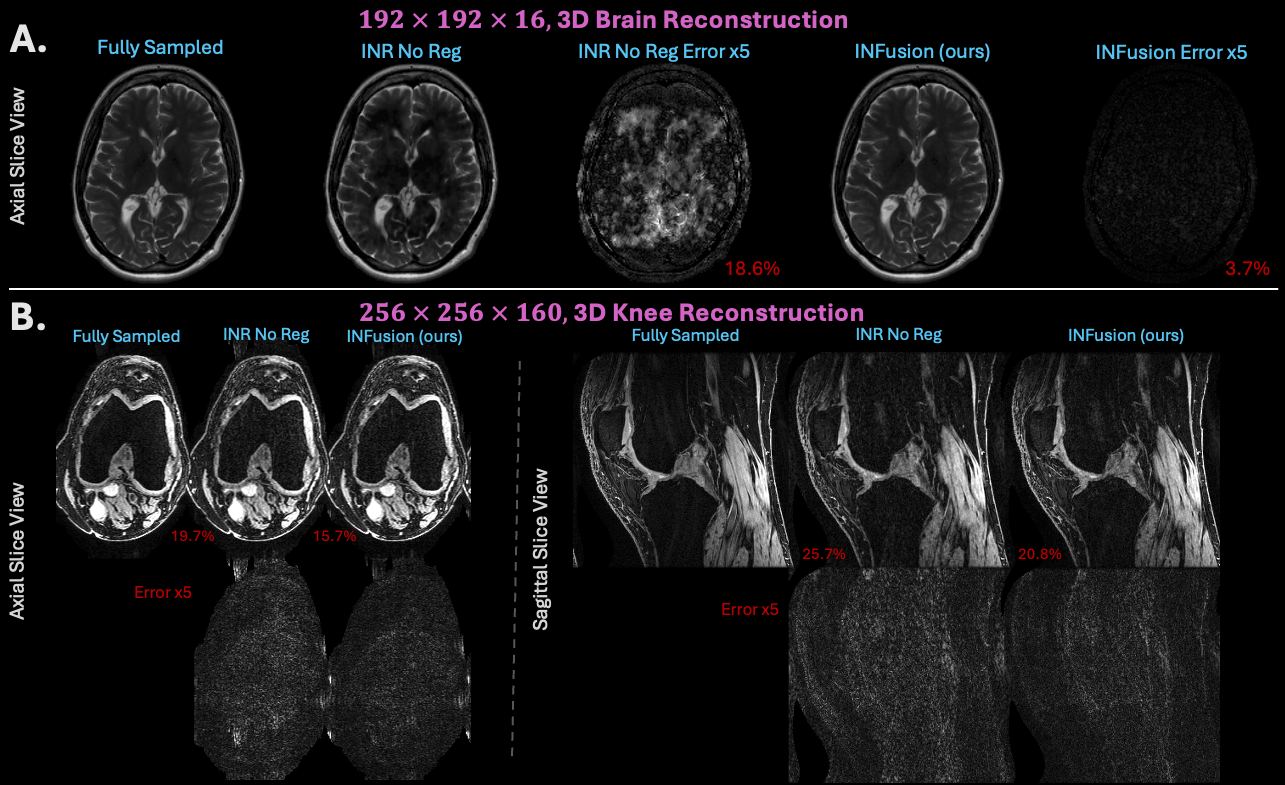}
    \caption{(A.) INFusion outperforms no regularization on the modest, single-coil 3D dataset, but encoding all three spatial dimensions in the INR required prohibitively large GPU memory usage and compute. (B.) Applying diffusion regularization to random slices at each iteration and encoding only the x-y coordinates enables training of INRs with diffusion regularization on a realistically sized $256 \times 256 \times 80$ 3D k-space volume.}
    \label{fig:3dresults}
\end{figure*}

In Fig. \ref{fig:3dresults}, A, INFusion outperforms no regularization on the modest, single-coil 3D dataset, but encoding all three spatial dimensions in the INR required prohibitively large GPU memory usage and compute.  Fig. \ref{fig:3dresults}, B shows that our proposed approach of diffusion regularization with random slices and just encoding the $x-y$ dimensions in coordinates enables training of INRs on $256 \times 256 \times 80$ matrix size k-space.

\section{Discussion}
INFusion exploits the generative capabilities of diffusion models to regularize INR training from scan-specific, under-sampled MR measurements. INR-based accelerated MRI inverse problems solved with our proposed diffusion regularization yield higher fidelity images with reduced scan time. We show improved performance in 2D and enable application in realistic 3D settings where the diffusion models were also trained on a different orientation than the under-sampling. Full 3D diffusion models are challenging to train, due to computational burden and limited data availability, but INFusion enables regularization of 3D reconstruction problems using 2D diffusion models. In addition, the 2D diffusion models can be trained on data from any 2D orientation that is available. 

Our implementation takes approximately 15 minutes to reconstruct a single 2D slice on a Nvidia A100 GPU, so clinical adoption requires a more efficient implementation. Future research work will investigate the effectiveness of INRs as an image representation and explore the trade-offs between using diffusion models as regularization versus posterior sampling directly on the voxel grid. 

\section*{Acknowledgment}
Support in part by NSF CCF-2239687 (CAREER), NSF IFML 2019844, and JCCO postdoctoral fellowship.

\bibliographystyle{IEEEbib}
\bibliography{strings,refs}

\begin{thebibliography}{10}

\bibitem{nishimura}
Dwight Nishimura,
\newblock {\em Principles of Magnetic Resonance Imaging},
\newblock Stanford University, 2010.

\bibitem{grappa}
Mark Griswold, Peter Jakob, Robin Heidemann, Mathias Nittka, Vladimir Jellus,
  Jianmin Wang, Berthold Kiefer, and Axel Haase,
\newblock ``Generalized autocalibrating partially parallel acquisitions
  (grappa),''
\newblock {\em Magnetic Resonance in Medicine}, 2002.

\bibitem{sense}
klaas pruessmann, markus weiger, markus scheidegger, and peter boesiger,
\newblock ``sense: sensitivity encoding for fast mri,''
\newblock {\em Magnetic Resonance in Medicine}, 1999.

\bibitem{espirit}
Martin Uecker, Peng Lai, Mark~J. Murphy, Patrick Virtue, Michael Elad, John~M.
  Pauly, Shreyas~S. Vasanawala, and Michael Lustig,
\newblock ``Espirit—an eigenvalue approach to autocalibrating parallel mri:
  Where sense meets grappa,''
\newblock {\em Magnetic Resonance in Medicine}, 2013.

\bibitem{lustig-cs}
Michael Lustig, David Donoho, and John~M. Pauly,
\newblock ``Sparse mri: The application of compressed sensing for rapid mr
  imaging,''
\newblock 2007.

\bibitem{modl}
Hemant~K Aggarwal, Merry~P Mani, and Mathews Jacob,
\newblock ``Modl: Model based deep learning architecture for inverse
  problems,''
\newblock {\em IEEE Transactions on Medical Imaging}, 2018.

\bibitem{varnet}
Kerstin Hammernik, Teresa Klatzer, Erich Kobler, Michael~P. Recht, Daniel~K.
  Sodickson, Thomas Pock, and Florian Knoll,
\newblock ``Learning a variational network for reconstruction of accelerated
  mri data,''
\newblock {\em Magnetic Resonance in Medicine}, 2017.

\bibitem{ssdu}
Burhaneddin Yaman, Seyed Amir~Hossein Hosseini, Steen Moeller, Jutta Ellermann,
  Kâmil Uğurbil, and Mehmet Akçakaya,
\newblock ``Self-supervised learning of physics-guided reconstruction neural
  networks without fully sampled reference data,''
\newblock {\em Magnetic Resonance in Medicine}, 2020.

\bibitem{robustcs}
Ajil Jalal, Marius Arvinte, Giannis Daras, Eric Price, Alexandros~G Dimakis,
  and Jon Tamir,
\newblock ``Robust compressed sensing mri with deep generative priors,''
\newblock 2021.

\bibitem{energy}
Jyothi~Rikhab Chand and Mathews Jacob,
\newblock ``Multi-scale energy (muse) plug and play framework for inverse
  problems,''
\newblock {\em arXiv}, 2023.

\bibitem{baymri}
Guanxiong Luo, Moritz Blumenthal, Martin Heide, and Martin Uecker,
\newblock ``Bayesian mri reconstruction with joint uncertainty estimation using
  diffusion models,''
\newblock {\em Magnetic Resonance in Medicine}, 2023.

\bibitem{mlreview}
Reinhard Heckel, Mathews Jacob, Akshay Chaudhari, Or~Perlman, and Efrat
  Shimron,
\newblock ``Deep learning for accelerated and robust mri reconstruction,''
\newblock {\em Magnetic Resonance Materials in Physics, Biology and Medicine},
  2024.

\bibitem{siren}
Vincent Sitzmann, Julien Martel, Alexander Bergman, David Lindell, and Gordon
  Wetzstein,
\newblock ``Implicit neural representations with periodic activation
  functions,''
\newblock {\em Advances in Neural Information Processing Systems}, 2020.

\bibitem{fourierfeatures}
Matthew Tancik, Pratul~P. Srinivasan, Ben Mildenhall, Sara Fridovich-Keil,
  Nithin Raghavan, Utkarsh Singhal, Ravi Ramamoorthi, Jonathan~T. Barron, and
  Ren Ng,
\newblock ``Fourier features let networks learn high frequency functions in low
  dimensional domains,''
\newblock {\em Advances in Neural Information Processing Systems}, 2020.

\bibitem{imjsense}
Ruimin Feng, Qing Wu, Jie Feng, Huajun She, Chunlei Liu, Yuyao Zhang, and
  Hongjiang Wei,
\newblock ``Imjense: Scan-specific implicit representation for joint coil
  sensitivity and image estimation in parallel mri,''
\newblock {\em IEEE Transactions on Medical Imaging}, 2024.

\bibitem{nerp}
Liyue Shen, John Pauly, and Lei Xing,
\newblock ``Nerp: Implicit neural representation learning with prior embedding
  for sparsely sampled image reconstruction,''
\newblock {\em IEEE Transactions on Neural Networks and Learning systems},
  2022.

\bibitem{inrcardiacbin}
Wenqi Huang, Hongwei~Bran Li, Jiazhen Pan, Gastao Cruz, Daniel Rueckert, and
  Kerstin Hammernik,
\newblock ``Neural implicit k-space for binning-free non-cartesian cardiac mr
  imaging,''
\newblock {\em Information Processing in Medical Imaging}, 2023.

\bibitem{inrfouriermri}
Johannes~F. Kunz, Stefan Ruschke, and Reinhard Heckel,
\newblock ``Fourier-feature inputs for free-breathing cardiac mri
  reconstruction,''
\newblock {\em IEEE Transactions on Computational Imaging}, 2024.

\bibitem{nerfmri}
Tae~Jun Jang and Chang~Min Hyun,
\newblock ``Nerf solves undersampled mri reconstruction,''
\newblock {\em arXiv}, 2024.

\bibitem{reconfusion}
Rundi Wu, Ben Mildenhall, Philipp Henzler, Keunhong Park, Ruiqi Gao, Daniel
  Watson, Pratul~P. Srinivasan, Dor Verbin, Jonathan~T. Barron, Ben Poole, and
  Aleksander Holynski,
\newblock ``Reconfusion: 3d reconstruction with diffusion priors,''
\newblock {\em Conference on Computer Vision and Pattern Recognition}, 2024.

\bibitem{stochasticscore}
Yang Song, Jascha Sohl-Dickstein, Diederik~P Kingma, Abhishek Kumar, Stefano
  Ermon, and Ben Poole,
\newblock ``Score-based generative modeling through stochastic differential
  equations,''
\newblock {\em International Conference on Learning Representations}, 2020.

\bibitem{edm}
Tero Karras, Miika Aittala, Timo Aila, and Samuli Laine,
\newblock ``Elucidating the design space of diffusion-based generative
  models,''
\newblock {\em Advances in Neural Information Processing Systems}, 2022.

\bibitem{lpips}
Richard Zhang, Phillip Isola, Alexei~A. Efros, Eli Shechtman, and Oliver Wang,
\newblock ``The unreasonable effectiveness of deep features as a perceptual
  metric,''
\newblock {\em Conference on Computer Vision and Pattern Recognition}, 2018.

\bibitem{fastmri}
Florian Knoll, Jure Zbontar, and et. al.,
\newblock ``fastmri: A publicly available raw k-space and dicom dataset of knee
  images for accelerated mr image reconstruction using machine learning,''
\newblock {\em Radiology: Artificial Intelligence}, 2020.

\bibitem{skmtea}
Arjun~D Desai, Andrew~M Schmidt, Elka~B Rubin, Christopher~M Sandino,
  Marianne~S Black, Valentina Mazzoli, Kathryn~J Stevens, Robert Boutin,
  Christopher Ré, Garry~E Gold, Brian~A Hargreaves, and Akshay~S Chaudhari,
\newblock ``Skm-tea: A dataset for accelerated mri reconstruction with dense
  image labels for quantitative clinical evaluation,''
\newblock {\em Advances in Neural Information Processing Systems}, 2021.

\bibitem{bart}
Jonathan Tamir, Joseph Cheng, Martin Uecker, and Michael Lustig,
\newblock ``Generalized magnetic resonance image reconstruction using the
  berkeley advanced reconstruction toolbox,''
\newblock {\em ISMRM Workshop on Data Sampling and Image Reconstruction}, 2016.

\bibitem{csreview}
Jong~Chul Ye,
\newblock ``Compressed sensing mri: a review from signal processing
  perspective,''
\newblock {\em BMC Biomedical Engineering}, 2019.

\end{thebibliography}

\end{document}